# Covalent functionalization of HiPco single-walled carbon nanotubes: the differences in the oxidizing action of $H_2SO_4$ and $HNO_3$ at soft oxidation process

Xavier Devaux[a], Brigitte Vigolo[a], Edward McRae[a], Fabrice Valsaque[a], Naoual Allali[b][d][f], Victor Mamane[b][c], Yves Fort[b], Alexander V. Soldatov[d][e], Manuel Dossot[f] and Svetlana Yu. Tsareva[a]

**Abstract:** This work presents the results of a study of the evolution of HiPco single-walled carbon nanotubes (SWCNTs) during the oxidizing action of $H_2SO_4$ and $HNO_3$. The process conditions used have been chosen so as to avoid any strong damaging of the nanotube structure. The type and level of functionalization, the location of the grafted functions on the nanotube surface and the changes in morphological characteristics of the samples were examined using a wide and complementary range of analytical techniques. We propose an explanation for the differences in the oxidizing action of sulphuric and nitric acids. The combined results allow us to suggest the possible reaction mechanisms which occur on the nanotube surface.

[a] Dr. S.Yu. Tsareva, Dr. X. Devaux, Dr. B. Vigolo, Dr. E. McRae, Dr. F. Valsaque
Institut Jean Lamour, UMR 7198 CNRS-Université de Lorraine
Faculté des Sciences et Technologies, Boulevard des Aiguillettes, BP 70239, 54506 Vandœuvre-lès-Nancy, France
E-mail: svetlana.tsareva@univ-lorraine.fr
[b] N. Allali, Dr. V. Mamane, Prof. Y. Fort
Laboratoire de Structure et Réactivité des Systèmes Moléculaires Complexes, UMR 7565 CNRS-Université de Lorraine
Faculté des Sciences et Techniques, Boulevard des Aiguillettes, BP 70239, 54506 Vandœuvre-lès-Nancy, France
Institution
[c] Dr. V. Mamane
Institut de Chimie de Strasbourg, UMR 7177, Equipe LASYROC, Université de Strasbourg
1 rue Blaise Pascal BP 296 R8, 67008 Strasbourg, France
[d] N. Allali, Prof. A.V. Soldatov
Luleå University of Technology, Department of Engineering Sciences and Mathematics,
SE-97187, Luleå, Sweden
[e] Prof. A.V. Soldatov
Department of Physics
Harvard University
Cambridge MA 02138, USA
[f] N. Allali, Dr. M. Dossot
Laboratoire de Chimie Physique et Microbiologie pour l'Environnement, UMR 7564, CNRS-Université de Lorraine
405 rue de Vandœuvre, 54601 Villers-lès-Nancy, France

# 1. Introduction

Carbon nanotubes (CNTs) and perhaps especially single-walled CNTs (SWCNTs) are attractive due to their potential in a variety of applications such as electronics devices, composite materials, energy storage devices, sensors and in medicine [1-6]. One hindrance, however, to all these potential uses is that as-grown SWCNTs rarely feature the characteristics required for certain specific applications. To transform these ideas into real-world devices, one or several stages of chemical modification are required [1-7]. A covalent modification or a covalent attachment of chemical functional groups is one possible way to realize this. A grafting of different functional groups is easier on an active oxygen-containing group such as -OH or -COOH on the CNT surface than directly on the graphene-like CNT sidewall [7-8]. Such groups can be generated by an oxidative process [7-13] which could be via heating in air or in an $O_2$ atmosphere, or via an acid treatment. Two often-used acids are $H_2SO_4$ and $HNO_3$ as well as their mixtures in different ratios [7, 9, 11-12, 14-16]. The level of functionalization can be quite different according to the treatment conditions and the structural properties of the CNTs [17]. Even if it is considered that carboxylic groups are prevalent on the CNT surface after an oxidative process, in reality there is always a variety of different oxygen-containing groups with a ratio that can be quite different from one process to another [7, 10, 13-16]. The exact mechanism that defines the preferential destination sites of oxygen-containing groups as well as the ratio between these groups are open to question. CNTs are more reactive in the tips, in the curved regions and in the connecting regions between different bundles [18]. Structural defects promote oxidation and favor the formation of certain functional groups [10]. During the acid treatment used, the electrophilic reagent first attacks some C=C bonds producing hydroxyl groups which in turn can be converted into quinone groups, and finally to carboxylic acid groups [19]. These last groups can form dimers [19] which under certain conditions (thermal treatment, for example) may transform into anhydrides [19]. Because the formation of oxygen-containing functional groups is stepwise ($-C-H \rightarrow -C-OH \rightarrow -C=O \rightarrow -COOH \rightarrow \cdots$) and the reactivity of the CNT surface is very heterogeneous, there is always a variety of different oxygen-containing groups after an oxidative process. The abundance and the ratio between the different functional groups depend on the nature of the oxidant [20]. In the case of a strong oxidant or when the oxidative process is applied for a long time, there are more functional groups on the CNT surface and the carboxyl groups predominate. The use of a very soft oxidation process in which the acids react with already-existing defect sites without any further damaging of the CNT sidewall, is the way which will lead to understanding the difference in the oxidizing action of different oxidants. But it is a real challenge to define the ratio between the number of the different grafted groups and their localization on the CNT surface, especially in the case of a low level of functionalization.

To verify the presence and number of grafted functional groups, different analytical methods such as infrared spectroscopy [4, 8, 13-16], X-ray photoelectron spectroscopy (XPS) [8] and thermogravimetric analysis coupled with mass spectrometry (TGA-MS) [6, 12, 15, 21] are commonly used. This list could be complemented with Raman spectroscopy that allows determining the kinds of SWCNTs most affected during treatment [4, 6, 14-17, 21]. Transmission electron microscopy (TEM) and scanning transmission electron microscopy (STEM) allow determining the level of damaging of CNTs as well as the evolution of sample morphology [4, 6, 8-9, 14-15, 17, 21]. In fact, there is no single method of analysis that gives complete information about the level of functionalization, the ratio between different grafted groups and their localization or the degree of sample damaging due to the process used. Only a combination of complementary methods can allow obtaining complete information that can shed more light on the interactions of CNT with different oxidants.

In this study we have compared the oxidizing action of $H_2SO_4$ and $HNO_3$. The concentration of the acids and the choice of the other process conditions used have been made so as to carry out "soft" oxidation in order to avoid any strong damaging of the CNT structure. HiPco Super Pure Tubes™ were used because their purity level is high which simplifies data interpretation since it is known that the acids can react with the impurities such as amorphous carbon, hollow carbon shells or catalyst particles.

We have used a whole battery of analytical techniques including TEM, STEM, Energy-dispersive X-ray spectroscopy (EDS), TGA-MS, Diffuse Reflectance Infrared Fourier Transform (DRIFT) spectroscopy, rare gas adsorption volumetry and Raman spectroscopy which collectively allow obtaining complete information about the type and level of functionalization, the distribution of the different grafted functional groups on the CNT surface and the changes in morphological characteristics of the samples. The combined results allow us to put forward an explanation for the difference in the oxidizing action of $H_2SO_4$ and $HNO_3$.

## 2. Results and discussion

### 2.1 *Raman scattering and DRIFT spectroscopy*

Fig. 1 presents resonance Raman spectra of pure-SWCNTs, oxH$_2$SO$_4$-SWCNTs and oxHNO$_3$-SWCNTs. It shows the radial breathing modes (RBM) at low wavenumbers (150–325 cm$^{-1}$) (in the inset), the graphene sheet-derived modes (the G band) around 1590 cm$^{-1}$ and the defect band (D-band) around 1300 cm$^{-1}$. This last is indicative of both the covalent defects on carbon nanotube side-walls and the carbonaceous impurities present in the sample [22]. Because the HiPco SuperPureTubes™ sample was purified by the manufacturer after synthesis, we attribute the D band mostly to these treatment-provoked defects within the tubes. The acid treatments seem to have preserved the CNT structural integrity and the relative D band intensity slightly decreases after both acid treatments. For the nanotubes resonant with the used laser wavelength, it is observed that the RBM band intensity for all metallic SWCNTs has decreased whereas that of the semiconducting band centered at 287 cm$^{-1}$ has slightly increased upon the acid treatment. Metallic SWCNTs are generally more sensitive to chemical treatments. Importantly, smaller diameter tubes of both types (metallic and semiconducting) are more influenced than the larger ones by functionalization as evidenced by the RBM profile. The G$^-$ component of the G-band exhibits the same trend as the RBM spectrum, i.e., there is a stronger impact of the H$_2$SO$_4$ treatment on the resonance conditions. Specifically, both the G$^-$ contribution from metallic nanotubes - the broad Breit-Wigner-Fano peak at about 1530 cm$^{-1}$ - and the G$^-$ peak from the semiconducting CNTs near 1550 cm$^{-1}$ decrease in intensity as shown on Fig. 1. Furthermore, functionalization of the semiconducting CNTs more strongly alters the resonance conditions and has higher impact on the G-band than on the RBM part of the spectrum.

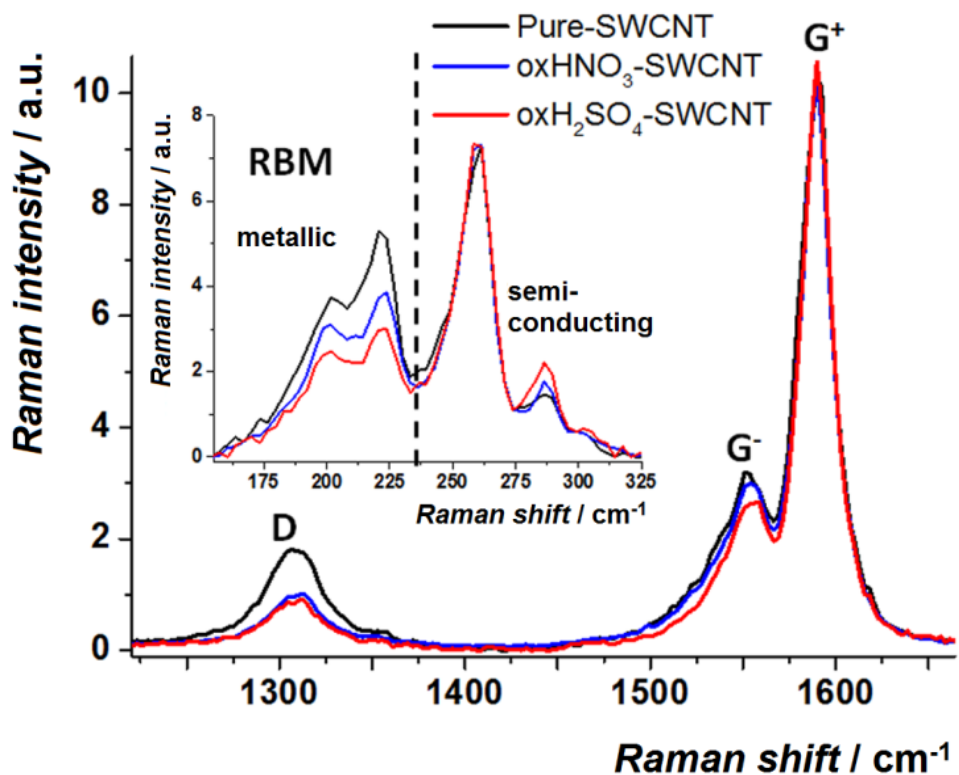

*Figure 1.* Raman spectra taken at 1.96 eV (633 nm) for the samples of pure-SWCNTs, oxH$_2$SO$_4$-SWCNTs and oxHNO$_3$-SWCNTs. The intensities have been scaled up to that of the G$^+$ peaks. In the inset at the top of the image: the RBM spectra taken for the same samples. The intensities of RBM have been scaled up to that of the 264cm$^{-1}$ band.

The interpretation of experimental results obtained from Raman spectra of SWCNT bundles is difficult. It is impossible to distinguish between the contribution to the D band from amorphous carbon and that coming from defective nanotubes. However, the defect formation in the SWCNT sample during the oxidizing process changes the intensity of all the Raman features and line shapes for the D, G and the G' Raman bands, and the lower frequency G$^-$ part of the G mode is more significantly reduced in intensity than the higher frequency G$^+$ part [23-24]. So, we can suggest that observed D band intensity slightly decrease accompanied with a decrease in the G$^-$ band intensity is a sign of the oxidation of both disordered carbon and nanotubes. Even if the number of functionalized SWCNTs after the acid treatments has increased, probably some amorphous carbon was eliminated from the samples during the oxidizing procedures. This results in a slight decrease of the D-band intensity of the two oxidized samples. Similar D-band behaviour has been described elsewhere [22].

An infrared spectrum of pure-SWCNTs is shown on Fig. 2. There is a very small broad band of hydroxyl groups as well as a very low intensity triplet at 2900 cm$^{-1}$. The latter is identified with C-H$_n$ functional groups [25] which probably come from the purification procedure that often involves an acid treatment to remove metal catalyst impurities. On the same figure there are the infrared spectra of oxH$_2$SO$_4$-SWCNT and oxHNO$_3$-SWCNT samples. The main features of these samples are assigned to two bands characteristic of carboxylic acid (at 1730 cm$^{-1}$) and quinone (at 1660 cm$^{-1}$) groups. A very broad band with a maximum near 3250 cm$^{-1}$ can be identified with -OH from the carboxylic acid and the phenol groups attached to SWCNTs. There is also a triplet around 2900 cm$^{-1}$ (2960, 2925 and 2850 cm$^{-1}$) of C-H$_n$ functional groups. A band at 1380 cm$^{-1}$ is associated with the -CH$_3$ groups. The intensity from all functionalities for the two oxidized samples is higher than what is seen in the spectrum of pure-SWCNTs. The band intensities of oxHNO$_3$-SWCNT sample are higher than those of the oxH$_2$SO$_4$-SWCNT sample. Since the quantity of the SWCNT samples diluted in KBr was approximately the same for all samples, the intensity of the bands can be associated with the density of functionalities.

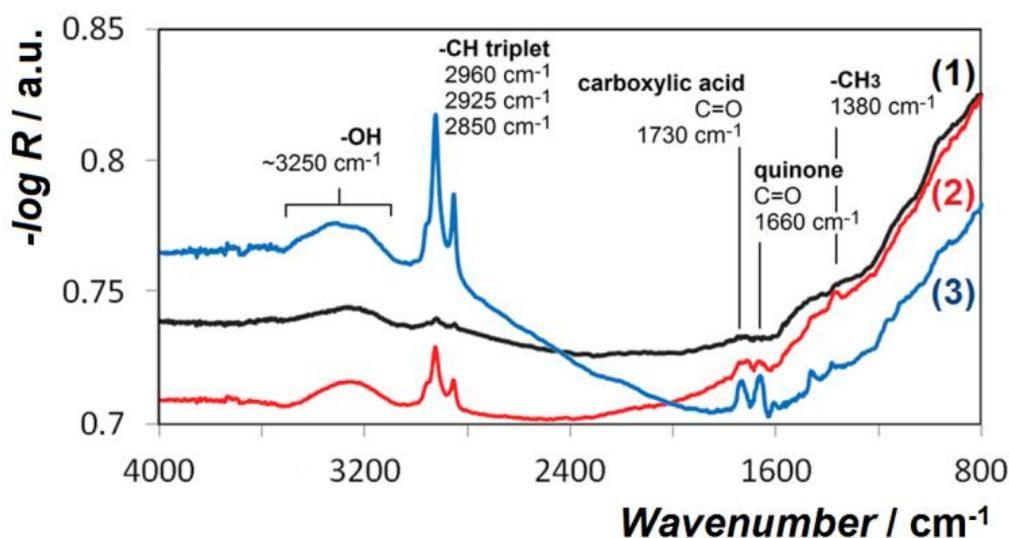

*Figure 2.* DRIFT spectra of pure-SWCNT (1), oxH$_2$SO$_4$-SWCNT (2) and oxHNO$_3$-SWCNT (3) samples.

We can thus conclude that in the case of HNO$_3$ treatment more functional groups were created than during the H$_2$SO$_4$ treatment. However, in both cases the oxidizing treatment was very soft. As was mentioned above, the formation of carboxylic acid groups passed through an intermediate stage of producing hydroxyl groups which can be converted into quinone and then to carboxylic groups [19]. So, the quinone groups may be considered as intermediates in the oxidation process [19]. The ratio between the peak heights at 1730 cm$^{-1}$ and 1660 cm$^{-1}$ corresponds to the ratio between carboxylic and quinone groups. As the oxidation proceeds, the ratio between the groups increases. For our two oxidized samples the ratio between carboxylic acid and quinone groups is approximately the same.

## 2.2 TGA-MS

In the thermal analyses, the weight loss of the different samples was recorded upon heating up to 1000°C under helium. The oxygen-containing functional groups, expected to be added through the two applied oxidative treatments, decompose leading to release of CO and/or $CO_2$ depending on the nature of the attached group [26]. Since the release of CO and $CO_2$ ends before 800°C, we give here the TGA-MS data only up to this temperature.

Fig. 3 (a) shows thermograms of pure-SWCNT and the two oxidized SWCNT samples. The pure-SWCNT undergoes a continuous weight loss as the temperature rises to reach a 15.2 wt.% loss at 800°C. The TG curves of the two oxidized samples are dissimilar. Below 700°C the weight loss of $oxH_2SO_4$-SWCNT is almost identical to that of pure-SWCNT but above 700°C there is an additional weight loss. For $oxHNO_3$-SWCNT, above 150°C the functions are eliminated with a more pronounced rate of weight loss. The treatment using $H_2SO_4$ has introduced fewer functions compared to that involving $HNO_3$. The total weight loss at 800°C reaches ~18 wt.% for $oxH_2SO_4$-SWCNTs and ~21 wt.% for $oxHNO_3$-SWCNTs which indicate a low level of functionalization for the two acids.

For pure-SWCNTs, the MS results (Fig. 3 (b)) show the emission of •$CH_3$ (m/z 15) and •$C_2H_2$ (m/z 26) fragments in the 100-600°C temperature range. The observed emission could come from the defect sites (the dangling bonds or hydrogen saturated carbon bonds) at the SWCNT surface. M/z 12 (carbon atoms) and m/z 28, corresponding to release of CO (m/z 12 being the fragment corresponding to carbon from CO), show a prominent increase in intensity around 660°C. It may come from phenol functions introduced by the purification treatment. That would mean that functional groups and defects are present on the surface of the purified SWCNTs before the oxidation treatments are carried out. This result correlates well with the IR spectroscopy data that have shown the presence -C-$H_n$ and -OH- containing groups in the pure-SWCNT sample.

Regarding the nature of the oxygen-containing functions introduced during acid treatment, MS signals recorded for m/z 44 ($CO_2$) (Fig. 3 (c)) show an increase with a bump around 140°C for ox$H_2SO_4$-SWCNT while the intensity of the signal for ox$HNO_3$-SWCNT slowly increases in the range 150-650°C. For both samples, $CO_2$ emission detected in the 150-650°C temperature range indicates the presence of carboxylic acid functions grafted at the SWCNT surface [26]. Loss of $CO_2$ is higher in this temperature range for ox$HNO_3$-SWCNT, in agreement with the greater weight loss for this sample compared to ox$H_2SO_4$-SWCNT (Fig. 3 (a)). Pure-SWCNTs emit a negligible quantity of $CO_2$ in this temperature range that can signify the presence of a few carboxylic groups on the SWCNT surface. This is probably why they were not detected by IR spectroscopy analysis.

Between 600° and 750°C, loss of CO is found for both samples (Fig. 3 (c)); it is related to the departure of phenol groups. For ox$H_2SO_4$-SWCNT, an additional peak around 700°C is observed. Moreover, unlike ox$HNO_3$-SWCNTs or pure-SWCNTs, for ox$H_2SO_4$-SWCNTs the peak for m/z 44 ($CO_2$) is also detected around 700°C (Fig. 3 (c)). Simultaneous emission of CO and $CO_2$ at this temperature is a characteristic of the anhydride groups [26]. The IR spectrum of ox$H_2SO_4$-SWCNT does not however show the presence of anhydride groups (Fig. 2). Feng et al. [18] showed that the carboxylic acid functionalities change drastically during thermal annealing. They proved using FTIR measurements that at a relatively low temperature (423 K) the anhydride may form from the coupling of two carboxylic groups with the loss of $H_2O$ (dehydrolysis). For the ox$H_2SO_4$-SWCNT sample we have verified that there is a signal for m/z 18, which is the main mass expected for water. There is a noticeable bump added to a continuously release during heating around 140°C (inset, Fig. 3 (c)). Therefore the excess water detected at 140°C is probably produced by the formation of anhydride functions from the coupling of two carboxylic acids. But in the case of ox$HNO_3$-SWCNTs in spite of having detected a greater number of carboxylic acid groups, we did not record the same dehydrolysis. Zhang et al. [19] showed that the efficiency of forming hydrogen bonds between -COOH groups is linked with their abundance, i.e. with their mutual proximity. So, we can suppose that for the two acid treatments the carboxylic acid functions are located differently. Indeed, in the ox$H_2SO_4$-SWCNT sample there are fewer grafted functions, but nevertheless they form dimers that transform into the anhydride functions during thermal treatment. This might be possible if the not numerous carboxylic acid groups are located densely, for example, at the end of the bundles.

A precise quantitative analysis is unfortunately not possible with such systems with a low level of functionalization containing a variety of functional groups that can react together and be released over quite large temperature range.

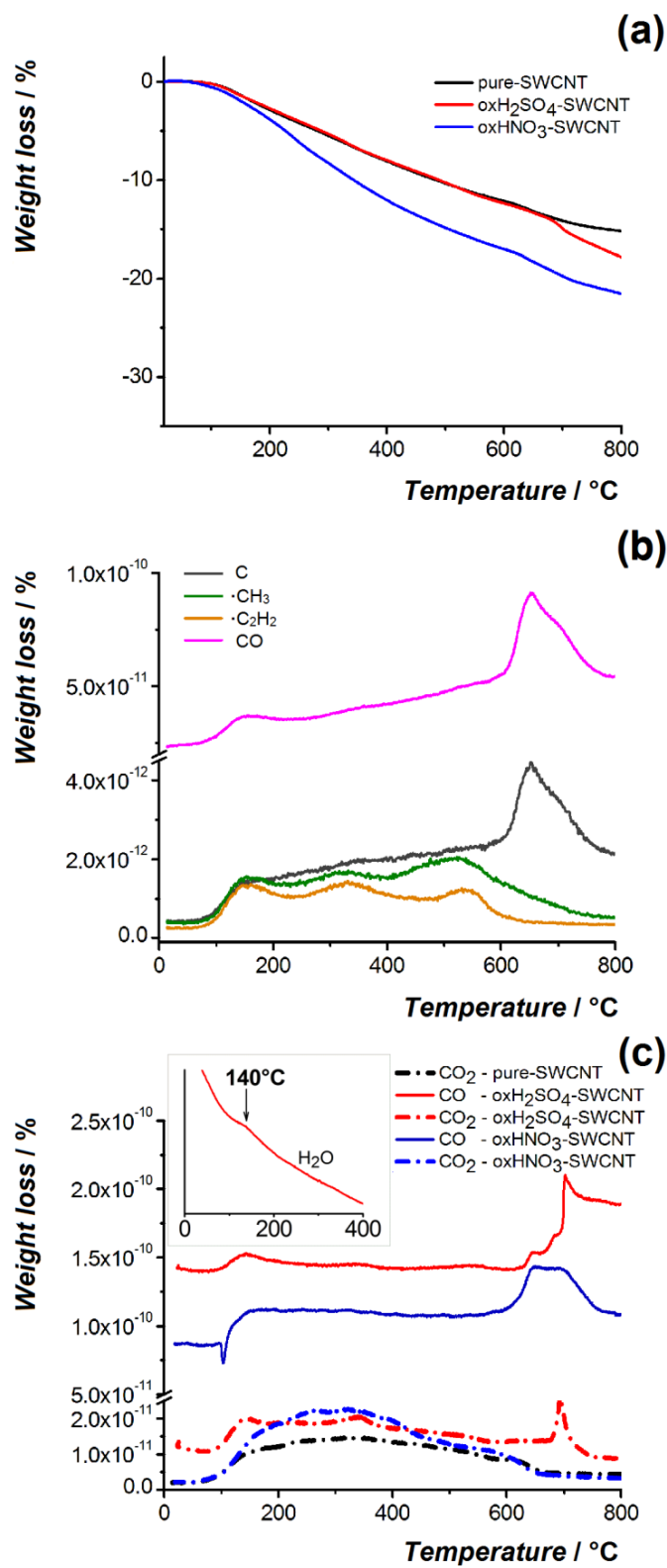

*Figure 3.* Thermograms taken under helium and the associated MS curves: (a) TGA curves of pure-SWCNT, oxH$_2$SO$_4$-SWCNT and oxHNO$_3$-SWCNT; (b) MS curves of C (m/z 12), •CH$_3$ (m/z 15), •C$_2$H$_2$ (m/z 26) and CO (m/z 28) for pure-SWCNTs; (c) MS curves of CO (m/z 28) and CO$_2$ (m/z 44) for pure-SWCNTs, oxH$_2$SO$_4$-SWCNT and oxHNO$_3$-SWCNT; insert: MS curve of H$_2$O (m/z 18) for oxH$_2$SO$_4$-SWCNT.

## 2.3 TEM and STEM examinations

As has been described elsewhere [27], detailed TEM/STEM examinations show that the pure-SWCNTs are combined into small bundles, comprising on average 25-30 nanotubes. There is also a significant number of isolated SWCNTs and very small bundles of 2 or 3 tubes. The bundles appear to be loosely bound and are some µm in length. The SWCNT diameters are heterogeneous, ~0.7-1.6 nm, with a consequential range of interstitial channel shapes and sizes, some of which are large. Triangular-shaped channels defined by three closely-packed tubes are observed, but there are other bigger channels defined by four tubes. The analysis of HRSTEM images shows that the distance between adjacent tubes is variable [27]. Most of the SWCNTs are closed. There are a few scattered metallic particles or carbon impurities. The carbon impurities comprise hollow carbon shells of 4–5 nm in diameter and small hollow spherical or ellipsoid structures of ~0.7–1.2 nm in size. These last hollow carbon "particles" are composed of one layer of carbon and might be fullerenes.

The detailed TEM-STEM examinations of ox$H_2SO_4$-SWCNT and ox$HNO_3$-SWCNT show that both acidic treatments do not significantly increase the number of opened SWCNTs and that the typical SWCNT lengths are similar to those of the pure-SWCNTs. A very small rise in the number of carbon impurities was observed in both cases. The morphologies of ox$H_2SO_4$-SWCNT and ox$HNO_3$-SWCNT were however different. As can be seen on Fig. 4 (a-b), the bundle structure of ox$H_2SO_4$-SWCNT is similar to that of pure-SWCNTs but in the case of ox$HNO_3$-SWCNT the morphology changes through a certain alignment of the bundles (Fig. 4 (c)). From DRIFT analysis we know that there are the same functional groups on the SWCNT bundle surface of the two samples. TGA-MS shows that the density of functionalities in the ox$HNO_3$-SWCNT is slightly higher than that of the ox$H_2SO_4$-SWCNT. TGA-MS also established a difference between the two oxidized samples regarding the thermal stability of the carboxylic acid groups that allow supposing a difference in the functional group localizations on the SWCNT surface. If we assume that in the case of ox$HNO_3$-SWCNT sample the carboxylic acids groups are uniformly dispersed on the side faces of the bundles, we can explain the alignment of the bundles by an electrostatic attraction between different bundles. This could be due to formation of the weak hydrogen bonds between functional groups on the CNT surface. The bundles in the ox$H_2SO_4$-SWCNT would then not be aligned because there are fewer -COOOH groups on the side faces of the bundles; they are probably located densely mainly at the ends of SWCNT bundles.

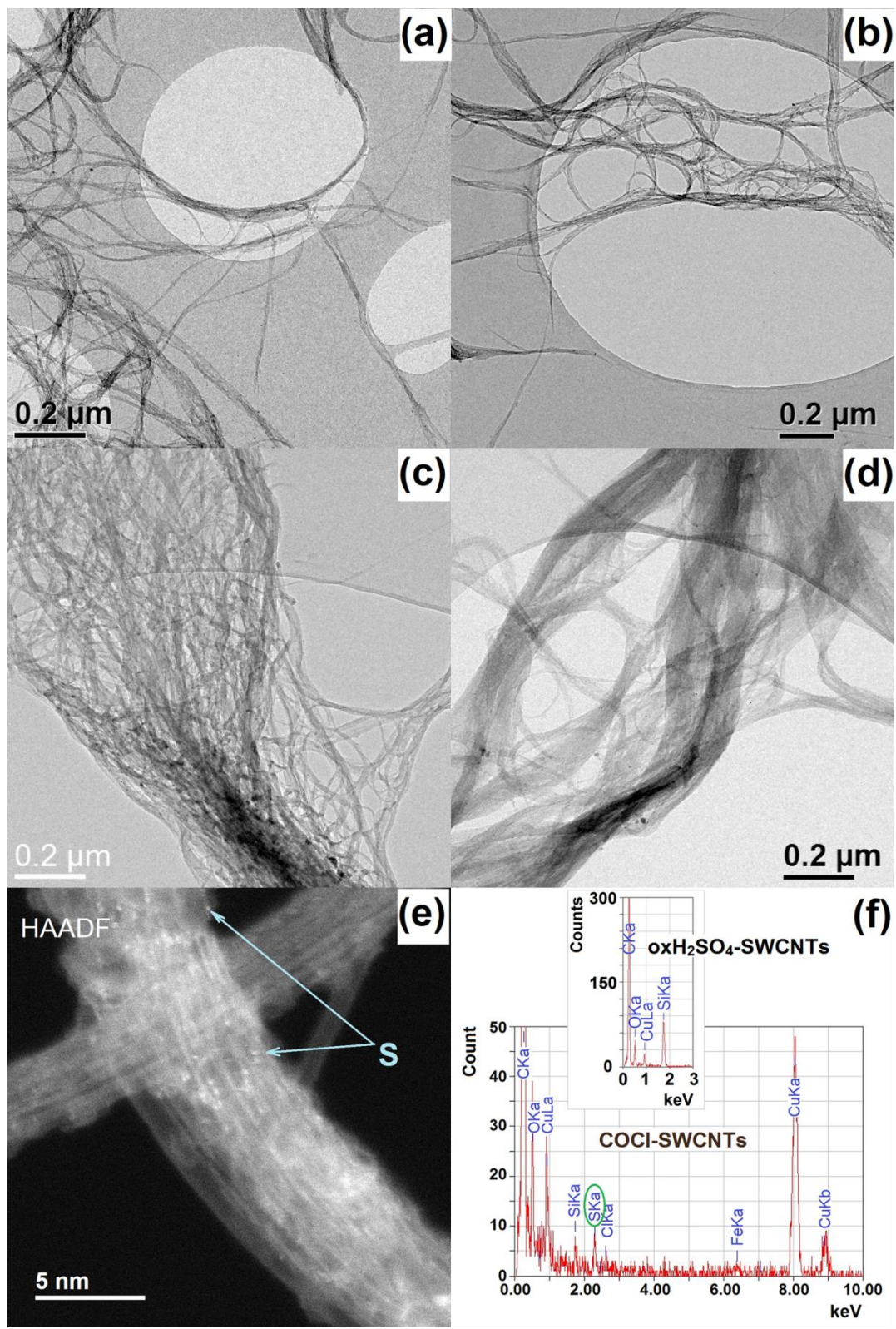

*Figure 4.* TEM (a-d), High-angle annular dark-field (HAADF) (e) images of (a) pure-SWCNTs, (b) $oxH_2SO_4$-SWCNTs, (c) $oxHNO_3$-SWCNTs, (d-e) COCl-SWCNTs. (f) EDS spectra of COCl-SWCNTs and $oxH_2SO_4$-SWCNTs (inset). The C, O, S, Si, Cl, Fe signals come from SWCNT samples, the copper signal comes from the sample grid.

Subsequent to the H$_2$SO$_4$ treatment, chlorination increases the bundle size (Fig. 4 (d)). On HAADF images (i.e., Fig. 4 (e)) on the SWCNT bundle surface one can see white points that correspond to atoms heavier than carbon. The EDS spectra show the presence of S and Cl (Fig. 4 (f)). The sulphur appeared on the SWCNT bundle surfaces only after the chlorination; it was no longer detected in the oxH$_2$SO$_4$-SWCNT. In the inset of Fig. 4 (f) there is an EDS spectrum taken from the side face of a bundle for oxH$_2$SO$_4$-SWCNT sample which shows the presence of Si (in the form of Si or SiO$_2$) on the surface of SWCNT bundles, but not sulphur. It should be noted that a Si signal was detected in all samples, even in pure-SWCNTs where it could come from the laboratory glassware used during purification or oxidation treatments. How the sulphur or chlorine might promote enlargement or consolidation of bundles will be discussed later.

**2.4 *Volumetric adsorption study***

Rare gas adsorption isotherms of SWCNT bundles generally have two inclined steps as the relative equilibrium pressure progressively increases [28-29]. The lower pressure step represents adsorption on higher binding energy sites. These are the grooves separating two adjacent outer tubes (G), the inner channels of the tubes (T) and the bundle interstitial channel (IC) sites. The higher pressure step corresponds to the exohedral adsorption on the external surface of the bundles (E). Being grafted on the SWCNT surface, functions can influence the adsorption site availability so this analytical technique can furnish information about changes which might have occurred during chemical treatments.

As has been described elsewhere [27], 77 K krypton adsorption isotherms on the pure-SWCNT sample used here have adsorptive dosing (AD) dependent characteristics of the low-pressure region (LPR) of the isotherm. This results from adsorption on the IC-sites that can be seen as subnanoscale pore channels with alternating enlargements (voids) and constrictions (necks) along the tube axes. An equilibrium pressure drop with increasing AD is noted using the increased adsorptive dosing (IAD) protocol that we have explained by the formation of metastable adsorbed phases inside the IC-sites or by intrapore blocking. In the measurements using the constant adsorptive dosing (CAD) protocol, the exact position of the first step is AD-dependent. It is located between about 0.0007 Pa and 0.05 Pa. Fig. 5 (a) presents the branches obtained using the CAD protocol for the four indicated doses of Kr. There is not a continuous step shift to the right as AD increases, the increasing AD influences the isotherm in a cyclic manner. The second step extends from near 0.3 Pa to about 3 Pa. Its position is independent of the protocol used and of the chosen AD value.

A soft acid treatment will graft oxygen-containing functions on the already-existing defect sites of a SWCNT sample without considerable damaging of the CNT sidewall. Kuznetsova *et al*. [30] showed that upon such an acid treatment, the grafted quinone or carboxylic acid groups block the entry ports into the individual nanotubes (T-sites). The bundle ends are the sites where the concentration of carboxylic acid functions after an acid treatment can be maximal [31]. They are also the entry ports to the ICs and opened SWCNTs. This means that if in the case of a pure-SWCNT sample, AD-dependent adsorption takes place in the IC-sites (and possibly T-sites of a few opened nanotubes), the subsequent acid treatment has blocked the entry ports to these sites leading to suppression of the branches, substeps and branch switching in the low-pressure region of the isotherm. Indeed, the 77 K adsorption curve of oxH$_2$SO$_4$-SWCNT has the "usual", AD-independent behavior (Fig. 5 (b)). There are no branches or equilibrium pressure drops with either the CAD or the IAD protocol. The first step extends from about 0.005 to about 0.02 Pa, i.e. it is much narrower than in the case of pure-SWCNTs. The reason underlying this change might be that the grafted functional groups like quinone or carboxylic acid groups are located at the ends of SWCNT bundles so they block the entry ports for Kr adsorption into the channel sites.

However, the adsorption curve of the oxHNO$_3$-SWCNT sample maintains an AD-dependent behavior. Fig. 5 (b) shows the branches in the LPR obtained for some different values of AD. This is a very surprising result. The data of IR spectroscopy and ATG-MS show that both acid treatments lead to introducing the same carboxylic acid groups on the SWCNT surface but in the case of HNO$_3$ the number of -COOH is slightly greater than for H$_2$SO$_4$. Since the tips of SWCNTs are among the most reactive sites of SWCNT bundles, one can expect here a localization of the same number of -C=O and -COOH groups as it is in oxH$_2$SO$_4$-SWCNT sample, resulting in blocking the entry ports into IC-sites for Kr adsorption. But this is not the case of the oxNHO$_3$-SWCNTs.

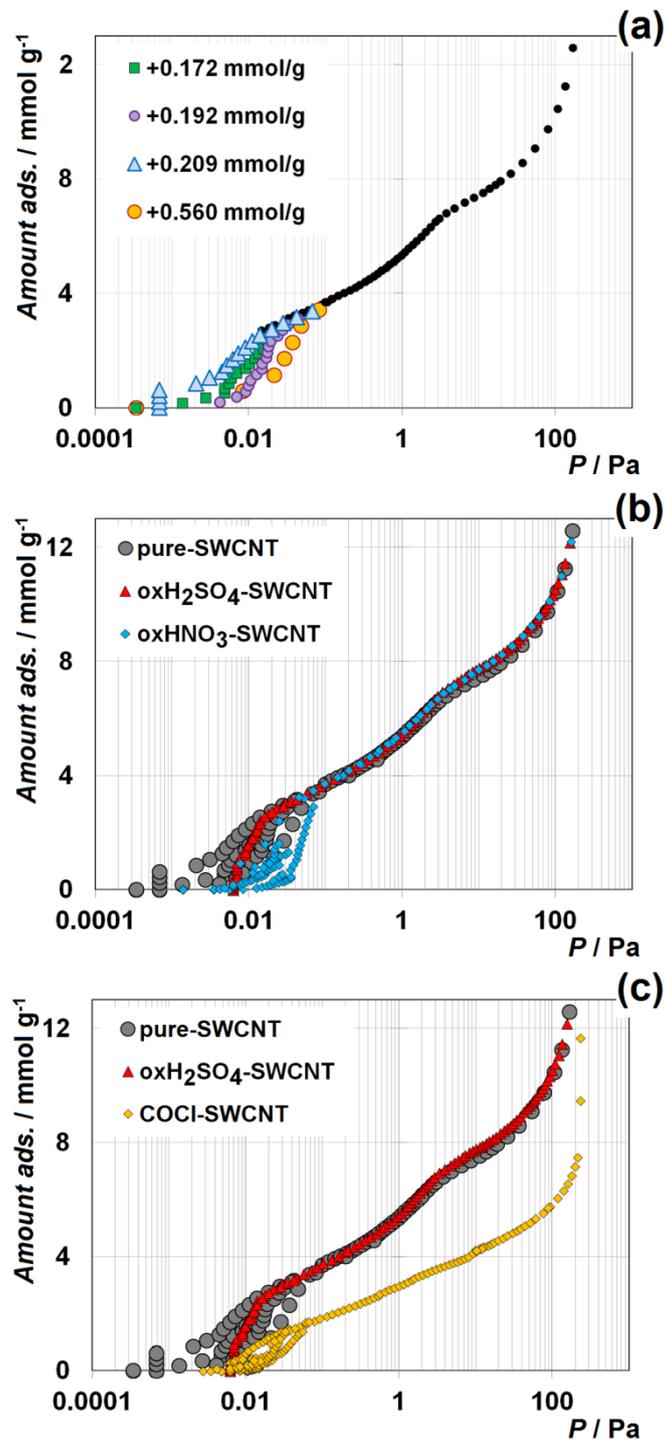

**Figure 5.** (a) 77 K isotherm of Kr adsorbed on pure-SWCNTs. In the LPR the branches obtained using the CAD protocol for the 4 indicated doses of Kr are presented. (b) 77 K isotherm of Kr adsorbed on pure-SWCNTs, oxH$_2$SO$_4$-SWCNT, oxHNO$_3$-SWCNT samples. (c) 77 K isotherm of Kr adsorbed on pure-SWCNTs, oxH$_2$SO$_4$-SWCNT, and COCl-SWCNTs. The adsorption curve of oxH$_2$SO$_4$-SWCNT has AD-independent behavior. For oxHNO$_3$-SWCNT and COCl-SWCNT samples the branches in the LPR were obtained using the CAD protocol for some different values of AD comparable with the indicated doses of Kr for pure-SWCNT sample.

So, the volumetric adsorption study has established a difference between the two oxidized samples regarding the number of the carboxylic acid groups on the tips of SWCNTs. For both oxidized samples the second part of the overall isotherm coincides with that of the pure-SWCNT sample (Fig. 5 (b)) implying that the number of functions on the surface is insufficient to influence the Kr adsorption on the external surface of the bundles.

Regarding the COCl-SWCNT sample, the 77 K Kr adsorption curve has AD-dependent behaviour (Fig. 5 (c)). This suggests that the entry ports to the IC-sites at the bundle ends that were closed by the $H_2SO_4$ treatment have been reopened by the chlorination treatment. The chlorination passes through the reaction SWCNT-COOH+$SOCl_2$→SWCNT-COCl [6]. The chlorocarbonyl groups stay in the same places that -COOH groups. Why in this case are the entry ports of the IC-sites opened for the Kr adsorption? One possible reason might be a breaking of the hydrogen bonds between two -COOH functions when then transform into -COCl groups. So, we can postulate that the grafted functional groups at the bundles ends such as quinone or carboxylic acid block the IC entry ports of the bundles only if they form dimers. If there are not enough adjacent functions to form dimers (the case of ox$HNO_3$-SWCNT sample) or the nature of the functions does not allow them to form hydrogen bonds (the case of -COCl groups), then they block only partially such entry ports through decreasing the size of the canal. So, the quantity of gas that can be adsorbed, then decreases.

The adsorption curve of COCl-SWCNT in the second step region is lower, i.e., the sample adsorbs less on the external sites. This could be explained by the bundle size increase. TEM micrographs (Fig. 4 (d)) show a rise in the bundle size so part of the initially external surface is "transformed" to an internal surface of the larger bundles, i.e. the total outer surface has diminished. On the HAADF image (Fig. 4 (e)) one can see on the SWCNT surface some atoms heavier than carbon. The EDS spectra show the presence of S and Cl (Fig. 4 (f)). The Cl signal arises from the chlorocarbonyl groups. They do not form the dimers. So, if the presence of Cl can be eliminated as an explanation, then S must be the reason for the bundle consolidation. In fact, during $H_2SO_4$ treatment on the SWCNT surface, not only carboxylic acid and quinone functions were created, but also phenol -OH groups (Fig. 2). If the bundles ends are occupied by the carboxylic acid and quinone functions, the phenol groups might be distributed over the side faces of the bundles. They do not form dimers like carboxylic acid and quinone functions explaining why they probably do not promote bundle alignment. But during the chlorination they can react with thionyl chloride forming sulfite ester functions linking two adjacent bundles: SWCNT-OH+SOCl$_2$→SWCNT-O-(S=O)-O-SWCNT.

### 2.5 *The possible reaction mechanisms that occur on the nanotube surface*

Let us summarize the obtained results. Firstly, all the used analytical techniques clearly show that the level of functionalization for the two oxidized samples is very low. In the case of oxHNO$_3$-SWCNTs, the number of oxygen-containing functionalities is slightly higher than in the oxH$_2$SO$_4$-SWCNTs sample. The oxygen-containing groups are the same in two samples: the phenol -OH, the quinone -C=O, the carboxylic acid -COOH groups. Nevertheless, the morphology of the two samples is different. TEM/STEM examinations show an oxHNO$_3$-SWCNT morphology change through alignment of the bundles. However, the morphology of oxH$_2$SO$_4$-SWCNT is similar to that of pure-SWCNT. We hypothesize that in the case of the HNO$_3$ treatment more functional groups such as quinone and carboxylic acids were uniformly located on the side faces of the bundles. The alignment might be due to an electrostatic attraction between different bundles owing to formation of the weak hydrogen bonds between functional groups on the CNT surface. The post-chlorination of oxH$_2$SO$_4$-SWCNT sample leads to increasing the average bundle size. STEM examination and EDS analysis show the presence of S on the SWCNT bundle surface which is introduced as a result of a reaction between thionyl chloride and phenol groups distributed on the SWCNT surface. As a result, the sulfite ester functions linking two adjacent bundles are formed. So, we can suggest that in the oxH$_2$SO$_4$-SWCNT sample on the side faces of the bundles there are phenol groups.

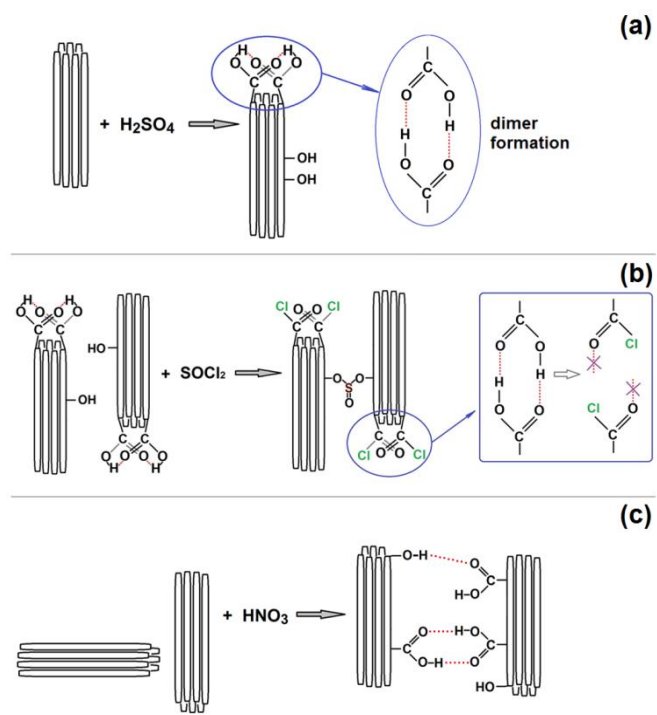

*Figure 6. Schematic illustration of the possible reaction mechanisms occurring on the nanotube during: (a) $H_2SO_4$ treatment; (b) the subsequent chlorination using $SOCl_2$; (c) $HNO_3$ treatment. Red dashed lines show the formation of the weak hydrogen bonds between different functions. The weak bonds between two groups located on two different bundles, are not strong enough to force the adjacent bundles to form any stable structure. But they can promote a bundles alignment noted by TEM/STEM examination.*

Since the formation of oxygen-containing functions is stepwise [20], the first hypothesis that occurs to us, is that the nitric acid is simply more reactive and is able to further oxidize the CNT sidewalls. It is known that the nanotubes ends are among the most reactive sites [10, 18]. So, they react the first with the acids. If that is case, then there should not be any difference in the number of -COOH groups at the nanotube ends of the two oxidized samples, and the only difference is in the number of functionalities along the nanotube length. This hypothesis can explain the difference in morphology between the two oxidized samples. But it cannot explain the difference regarding the results of ATG-MS or the volumetric adsorption study. It cannot explain, why the same carboxylic acids groups located at the ends of SWCNT bundles, transform into the anhydride functions during thermal treatment in the ox$H_2SO_4$-SWCNT sample and not in the ox$HNO_3$-SWCNTs. Furthermore it cannot explain, why -COOH groups block the entry ports to the IC and T-sites for Kr adsorption in the ox$H_2SO_4$-SWCNT sample and not in the ox$HNO_3$-SWCNT.

But if we hypothesize that the number of -COOH functional groups at the ends of the SWCNT bundles in the two oxidized samples is different, we can explain all obtained results including those of ATG-MS and volumetry. -COOH groups densely located at the end of bundles in oxH$_2$SO$_4$-SWCNT sample, form the dimers between two adjacent groups. The dimers block the entry ports to the interstitial channels and to the opened tubes for Kr adsorption. During the thermal treatment they transform into the anhydride functions. In the oxHNO$_3$-SWCNT sample there are few -COOH groups (or maybe even none) at the tips of nanotubes, i.e. there are no dimers of -COOH. For this reason the IC and T-sites stay opened for Kr adsorption. For the same reason there is no transformation of the dimers of -COOH groups into the anhydride groups during the thermal treatment.

Fig. 6 presents the possible reaction mechanisms occurring on the nanotubes. Of course, the presented images are only schematic; they illustrate the localization of the reactions taking place and the functionality organization on the CNT surfaces. These include the dimer formation between two adjacent -COOH groups at the bundle ends (a), the formation of "bridge" functionalities linking two adjacent bundles (b) and the bundle alignment due to an electrostatic attraction owing to formation of the weak hydrogen bonds between functional groups on the CNT surface (c).

We do not know the true reason of this difference in the oxidizing action of two acids. We can only suppose that the reason might be in a different oxidizing force of two acids. Probably, the nitric acid is more effective, i.e., stronger and for the same time of interaction with the CNTs, more oxygen-containing functionalities are formed on the nanotube surface than with H$_2$SO$_4$. Because it is more reactive, it probably reacts immediately with all the reactive sites. But it still remains unclear, why during the HNO$_3$ treatment the transformation of phenol groups to carboxylic acids at the nanotube tips was not realized in contrast to the side faces of the bundles. We believe that this difference in the number of -COOH groups at the bundle ends in the two oxidized sample is characteristic of a soft oxidation process. A particularity of the current study is in realization of a very soft oxidation. This is probably why we have observed this phenomenon. Using a wide and complementary range of analytical technique has allowed determining the difference between some properties of oxidized samples resulting from different numbers of -COOH groups at the nanotube ends.

## 3. Summary and conclusions

In this work we have presented the results of a study on the covalent functionalization of HiPco SWCNTs. The analytical techniques including TEM/STEM, EDS, TGA-MS, Raman scattering, DRIFT spectroscopy, rare gas volumetric adsorption have allowed obtaining much information about the level of functionalization, the type of functional groups and their location on the nanotube surface.

Both $H_2SO_4$ and $HNO_3$ treatments lead to introducing the same functions on the SWCNT bundle surfaces, but the number of -COOH groups at the end of SWCNT bundles is different, resulting in significant property differences. In the $oxH_2SO_4$-SWCNT sample, quinone or carboxylic acid groups grafted at the bundle ends can block the entry ports into the inner channels of opened tubes or into the interstitial channels of bundles via dimer formation. In the case of the $oxHNO_3$-SWCNT sample, the -COOH functions well dispersed on the SWCNT surface can form weak bonds either with -COOH, or with other groups like -OH or quinone =O located on adjacent bundles. This results in an alignment of the bundles.

Subsequent chlorination of the $oxH_2SO_4$-SWCNT sample transforms -COOH groups into the chlorocarbonyl groups COCl-. -OH groups distributed over the SWCNT surface, which react with thionyl chloride with formation of sulfite ester functions that can link the adjacent bundles. This results in an increase of the bundle size.

**Experimental Section**

1. *Samples*

The SWCNTs used in this study are HiPco SuperPureTubes™ produced by NanoIntegris. According to the manufacturer, the diameter range is 0.8-1.2 nm with individual CNT lengths up to 1 μm; they contain <5% of metal catalyst impurities. The SWCNTs are bundled.

**Oxidation of SWCNTs using acids** 10 mg of HiPco SuperPureTubes™ were placed in a 10 mL vial and 3 mL of concentrated $HNO_3$ (70% w/w which corresponds to 15 M) or 2.5 M $H_2SO_4$ were added. The vial was closed with a Teflon cap and irradiated under microwaves at 50°C for 20 min (15 psi, 100% power with a 300 WCEM Discover). After cooling, the SWCNTs were submitted to three centrifugation/decanting cycles in deionized water (10 mL for each cycle). The SWCNT acid suspensions were neutralized by adding a sodium hydroxide solution (2M, 4 mL) and washed extensively with water using sonication to eliminate the salts of sodium. The final rinsing was performed with ethanol to remove all traces of water. The Ox-SWCNTs were dried overnight in an oven at 80 °C and stored in closed vials at room temperature.

**Treatment of SWCNTs with $SOCl_2$** Thionyl chloride (10 mL) was added to ox-SWCNT (10 mg) and the mixture was refluxed at 70°C for 24 h under argon atmosphere. The thionyl chloride was removed under vacuum at ambient temperature then the SWCNTs were washed several times with anhydrous tetrahydrofuran (THF) using three sonication/filtration cycles (20 mL THF for each cycle). The COCl-SWCNTs were dried under vacuum for 5 h. The COCl-SWCNTs can be stored under argon at -20°C for several days.

Due to the instability of chlorocarbonyl functional groups at ambient temperature in air, some methods of analysis that would have required long manipulations under these conditions, were not carried out.

Hereafter for HiPco SuperPureTubes™ we will use the term "pure-SWCNT", for SWCNTs oxidized with $H_2SO_4$ - "$oxH_2SO_4$-SWCNT", for those oxidized with $HNO_3$ - "$oxHNO_3$-SWCNT", and finally for $oxH_2SO_4$-SWCNT chlorinated with $SOCl_2$ the term "COCl-SWCNT".

## 2. *Raman spectroscopy*

Raman scattering experiments using a He-Ne laser excitation (633 nm wavelength) were done on a Witec CRM200 confocal Raman microscope equipped with an electrically-cooled CCD detector, a grating with 1800 grooves per mm (spectral resolution 1 $cm^{-1}$), and a long working distance x50 objective with a numerical aperture of 0.55. CNTs were dispersed in tetrahydrofuran (THF) (0.1 mg of CNTs / mL) using sodium deoxycholate (DOC) as a surfactant (5% in weight). A tip sonication method was used to make the suspension (90 W for 30 min) which was then spin-coated onto a glass slide. The laser irradiance was kept below 1 kW $cm^{-2}$ to avoid any laser heating damage of the samples. In order to strengthen the statistical meaning of the Raman data, more than 40 spectra were collected at various locations on the sample, and we report below the average spectra for each sample.

### 3. *Diffuse Reflectance Infrared Fourier Transform (DRIFT) Spectroscopy*

Infrared powder diffuse reflectance data were collected with a Harrick Praying Mantis™ attachment and a high temperature reaction chamber that allowed analyzing the sample up to 100°C under a nitrogen flow. Spectroscopic grade KBr was used as background. The diffuse reflectance $R_s$ of the sample and $R_r$ of potassium bromide, used as a non-absorbing reference powder, were measured under the same conditions. The sample was prepared by mixing the SWCNT with KBr (SWCNT mass fraction of 0.001 in KBr), without compaction. The reflectance is defined as $R=R_s/R_r$. The spectra are shown on a pseudo-absorbance (-logR) scale.

### 4. *Thermogravimetric analysis coupled with mass spectrometry (TGA-MS)*

A Setaram Setsys Evolution 1750 Thermal Gravimetric Analyzer coupled with a Pfeiffer GSD 301C Vacuum OmniStar mass spectrometer allowed studying the nature of the products formed by detachment of the groups from the SWCNT surface and their release into the TGA system. About 3-4 mg of raw or functionalized sample were placed in an alumina crucible in the TGA chamber and the temperature was raised at a rate of 3°C/min from room temperature to 1000 °C under a helium (Air Liquide Alphagaz 2) flux of 20 mL/min. The parameters used for the mass spectrometer apparatus ensure that most of the species undergo single ionization, meaning that the charge $z$ = 1 for the detected mass-to-charge ratio *m/z*; *m/z* or mass will thus be indifferently employed in the text. TGA-MS characterizations are performed on at least two samples for any given batch in order to guarantee reproducible results.

### 5. *Electron microscopy*

Samples were dispersed in absolute ethanol then deposited on a holey carbon film supported by a 300 mesh copper grid. Some of the high-resolution transmission electron microscopy

(HRTEM) observations were done using a Philips CM200 operating at 200 kV. The others, HRTEM and scanning transmission electron microscopy (STEM) images, were done with a JEOL ARM 200F equipped with a cold field-emission gun, a probe Cs-corrector (correction of the spherical aberration) and with JEOL SD30Gv detector for EDS spectroscopy. The microscope was operated at 80 kV to avoid damaging the SWCNTs. High-angle annular dark-field (HAADF) STEM images were obtained for collection semi-angles of 45-180 mrad, and a pixel time of 40 µs (1024x1024 pixels).

## 6. *Volumetric measurements*

The physisorption studies were carried out using an apparatus with two capacitive pressure gauges allowing measurement between $10^{-3}$ and 1100 Pa (*i.e.*, over the full range of the isotherms) with a resolution of 0.003% of full scale and a precision of 0.5% (in the range $10^{-3}$-1 Pa) and of 0.2% (in the range 1-1100 Pa) of the reading. The raw isotherms were corrected for the thermal transpiration effect [32]. A liquid $N_2$ bath temperature stabilization method was used to maintain a constant temperature (77 K) of the sample cell containing the adsorbent under study. The other part of the apparatus (including pressure gauges, inlet manifold, gas reservoir, etc, except the pumps) isolated from the cell by a valve, was stabilized at 30.0±0.5°C. Krypton was first introduced into this latter part and then into the adsorption cell by opening the valve. Upon adsorption, the cell pressure thus decreases and reaches a limiting value at equilibrium. Equilibration times ranged from 0.3 to 1 h per dose and were chosen experimentally. Our experimental setup was optimized so that small mass samples - but with large specific surface areas - could indeed be examined. The samples (typically 10 mg) were initially outgassed outside the adsorption chamber for 7 days at 100°C to a pressure lower than $10^{-4}$ Pa. After installation in the adsorption set-up, they were additionally outgassed for 1 day at 100°C to a pressure lower than $5.10^{-5}$ Pa. Further isotherm measurements always started with outgassing of the sample at room temperature (20°C) for 16 hours to a pressure of $< 5.10^{-5}$ Pa. Krypton ($\geq$99.998%, Fluka Analytical) was purified by pumping on the condensed phase inside the apparatus. The experiments were carried out with use of two adsorptive dosing (AD) protocols. These are designated IAD (increased adsorptive dosing) if the dose is increased from one injection to another or CAD (constant adsorptive dosing) if the dose is constant.

**Acknowledgements**

Acknowledgment is made to the French Agence Nationale de la Recherche (ANR) (grant ANR-10-BLANC-0819-01-SPRINT) and the Région Lorraine (grant 30031172) for financial support. The authors thank Prof. C. Carteret and Dr. S. Fontana for fruitful discussions and Dr. J.-F. Marêché for important technical assistance.